\documentclass[11pt]{article}
\usepackage[paper=a4paper,dvips,top=2.5cm,left=1.5cm,right=1.5cm,bottom=2.6cm,footskip=2.2cm,voffset=0.0cm, marginparwidth=18mm]{geometry}
\usepackage{amsfonts,amsmath,amsthm,amssymb}

\numberwithin{equation}{section} 
\usepackage[svgnames]{xcolor}
\usepackage{fix-cm}
\usepackage{sectsty}
\usepackage{fancyhdr}
\usepackage{lastpage}
\usepackage{amsmath}
\usepackage{graphicx}
\usepackage{url}
\usepackage{enumerate}
\usepackage{paralist}
\usepackage{multicol}
\usepackage{color}  
\usepackage{float}  
\usepackage{epsfig}
\usepackage{subfigure}
\usepackage{hyperref}   
\hypersetup{colorlinks=true,linkcolor=beamer@PRD, citecolor=beamer@PRD}
\usepackage{authblk}  
\usepackage{cite}  
\usepackage{todonotes}
\usepackage{ulem} 

\renewcommand\eqref[1]{\textcolor{beamer@PRD}{(}\ref{#1}\textcolor{beamer@PRD}{)}}

\definecolor{beamer@PRD}{RGB}{46,48,146}

\begin{document}
\title{\textbf{Circuit Complexity for Coherent-Thermal States in  Bosonic String Theory}}
\author{\small{Arshid Shabir$^1$, Sanjib Dey$^2$, Salman Sajad Wani$^3$,  Suhail Lone$^4$,   Seemin Rubab$^5$,  Mir Faizal$^{6}$}
\\
\textit{\small $^{1, 5}$Department of Physics,  National Institute of Technology, Srinagar
Srinagar, Jammu and  Kashmir 190006, India}
\\
\textit{\small $^{2}$Department of Physics, BITS-Pilani, K. K. Birla Goa Campus, Zuarinagar, Goa 403726, India}
\\
\textit{\small $^{3, 6}$Canadian Quantum Research Center, 204-3002, 32 Ave Vernon, BC V1T 2L7, Canada}
\\
\textit{\small $^{3}$Department of Physics and Engineering, Istanbul, Technical University, 34469, Istanbul, Turkey}
\\
\textit{\small $^4$Department of Physics and Astrophysics, University of Delhi, Delhi 110007, India}
\\
\textit{\small $^6$Department of Physics and Astronomy,
University of Lethbridge,} \\ \textit{\small Lethbridge, AB T1K 3M4, Canada}
\\
\textit{\small $^6$Irving K. Barber School of Arts and Sciences, University of British Columbia}\\
\textit{\small Okanagan Campus, Kelowna, V1V1V7, Canada}
}
\date{}
\maketitle
\begin{abstract}
In this paper, we first construct thermofield double states for bosonic string theory  in the light-cone gauge. We then obtain a coherent-thermal string state and a thermal-coherent string state. We use  the  covariance matrix approach to calculate the circuit complexity of coherent-thermal string states. In this approach,   we generate the optimal geodesics by a horizontal string generator, and  then  obtain  the circuit  complexity using the  length of the minimal geodesics in the group manifold. 
\end{abstract}	 
\section{Introduction} \label{sec1}
\addtolength{\footskip}{-0.2cm} 

The circuit complexity of a quantum system indicates the minimum number of gates in the form of unitary operators necessary to reach a specific target from a particular reference state. Such a study has become important in AdS/CFT correspondence \cite{1} because this duality leads to the possibility of holographically obtaining the circuit complexity of the boundary theory from the bulk AdS geometry \cite{3, 5}. The holographic complexity of the boundary theory has been related to bulk AdS geometry using two different proposals. In the first proposal, the holographic complexity of the boundary theory is related to the codimension $-1$ maximal space-like surface in the bulk AdS geometry \cite{4}. In the second proposal, the holographic complexity of the boundary theory is obtained by evaluating the gravitational action on a Wheeler-de Witt patch of the AdS geometry \cite{6,7}. These conjectures have been studied in the circuit complexity of boundary theories dual to  different AdS solutions    \cite{8,9, 11,12,13,14}. They have also been used to analyze the behavior of holographic complexity for such interesting boundary theories  \cite{15,16,18,19,20}.  The circuit  complexity has become important due to its application in black hole physics \cite{2, 21,22,23,24,25,26}. It has been suggested that holographic complexity can be used to understand the black hole information paradox  \cite{27,28,29,30,31,32,33,34}.  
 
In quantum computation, it is crucial to find the minimum number of quantum gates needed to construct a unitary operator to perform a computation efficiently \cite{k0, k1}. Thus, complexity    quantifies the difficulty in carrying out a specific computational task. So,  in the context of quantum computing,  we can find a unitary operation whose role is to map an input quantum state for a specific number of qubits to an output quantum state with the same number of qubits \cite{41,42,43}. We can construct a unitary operator in  a circuit space using some elementary gates. We can use multiple methods to construct such an operator  to a specific accuracy. While constructing the desired unitary operation  up to some tolerance, the minimum number of elementary gates required to construct such a unitary operator gives the circuit complexity of the unitary operator.

We would like to point out the  difference between circuit complexity for states and circuit complexity for unitaries.  The circuit complexity for unitaries  represents the  
size of the system and is  measured using  the number of gates, of the smallest circuit that 
effects the unitary. The circuit complexity for a  state is defined as the size of the smallest circuit that produces that state from another  state \cite{open linear}. Thus, to define the circuit complexity for a state, the main step would be to identify the reference and target state. The minimum number of required gates to go from the reference to the target state is then defined as the circuit complexity, which can also be obtained geometrically from the space of circuits using the Nielsen formalism  \cite{k2}.  A Riemannian metric can be constructed in such a space, and then finding the   length of the shortest geodesic in this space of circuits is equivalent to finding the circuit complexity   \cite{k4}.
One can then analyze and select the optimal circuit out of the infinite number of possible circuits connecting the reference state and the final target state.  

Thus, the geometric idea of finding a minimal length in a  Riemannian metric  of qubits   \cite{k4} can be generalized to field theory, where we can find such an optimal circuit. To generalize the concept of circuit complexity to a free field theory, we observe that the  number of degrees of freedom in a field theory is infinite.
The circuit complexity of the ground state of a free scalar field theory has been obtained  \cite{44}. It has been done using the  Nielsen formalism \cite{k4} by relating the circuit complexity of a quantum field theory to the   length of the shortest geodesic in the space of circuits. Complexity for  free fermionic theories has also been studied using this geometric  approach     \cite{46,47}.  Here the Nielsen formalism \cite{k4}, is again used to relate  the length of the shortest geodesic in the space of circuits   to the circuit complexity of fermionic field theories.   The  Fubini-Study metric has also been used to obtain the circuit complexity of free field scalar theory \cite{48}.  
A  complementary approach to complexity in  the quantum field theory has been proposed, and this is based on the path-integral techniques \cite{52,53,54}.   The  complexity for strongly coupled large $N$ body systems has   been studied  holographically  using the Fubini-Study metric \cite{55,56}. The Nielsen formalism \cite{k4} has been used to obtain  circuit complexity for  interacting field theories   \cite{57}. The  integral transforms were used to analytically perform the   lattice sums for such interactive field theories to obtain their circuit complexity. 

To investigate  the thermal properties of a quantum system, it is possible to   construct a thermofield double TFD state from two copies of such a system \cite{tf, tf1}. 
This state has also been constructed for a free field theory.   The circuit complexity of thermal states in such theories has also been studied \cite{36,49,50,51}. This has been done by using a  target state and a reference state.   A composition of two copies of the  state  \cite{44,48} has been  used as the reference state  \cite{36}.  Two unentangled copies of the vacuum state have also been used    as the reference state  \cite{49,50}. It may be noted that  due to the difference in reference states, the circuit 
complexity  in these two different approaches is different.  The circuit complexity for Proca theory has been studied using Nielsen formalism \cite{nf1}. This was done by using the specific states of Proca theory, and it was observed that the circuit complexity for such states  has a logarithmic growth.  The  circuit complexity for a two-dimensional conformal field theory has also been constructed using  conformal symmetry transformations \cite{cft, cft1}.  

It is possible to study the circuit complexity for a coherent state by using  the covariance matrix approach  \cite{38}.
 The  circuit complexity of the general coherent state has also been constructed using the  covariance matrix approach \cite{a}.  The coherent-thermal (CT) state and thermal-coherent (TC) state are the two types of coherent states in a thermal system. Using the covariance matrix approach  \cite{36},   the circuit complexity for the coherent-thermal (CT) state was obtained \cite{a}.  This circuit complexity could  be read from the norm of the horizontal generator as the optimal geodesic. In this paper, we will use the  formalism \cite{36,   38, a } to analyze the circuit complexity for bosonic string theory. We will first construct thermofield double states for bosonic string theory in the light-cone gauge. Then we will obtain a coherent-thermal string state, and thermal-coherent string state, and  use coherent-thermal string state to calculate the circuit complexity. This will be  done by using the covariance matrix approach to calculate the circuit complexity of coherent-thermal string states. We will  generate  the optimal geodesics by a horizontal string generator in  the space of   circuits for  string states, and    use it  to calculate the circuit complexity of string states. This will be  done by calculating the length of the minimal geodesic in this space of circuits for  string states. In this paper, we will use the previous work done on the complexity of a simple harmonic oscillator \cite{a}, and the close analogy between the world-sheet modes of a bosonic string with a simple harmonic oscillator \cite{wani}  to obtain a complexity of a bosonic string theory. 
\section{TFD state of quantum bosonic strings}
\lhead{TFD state of quantum bosonic string}
\chead{}
\rhead{String State}
\addtolength{\voffset}{0.8cm} 
\addtolength{\footskip}{-0.7cm} 
In this section, we shall build the TFD state for the bosonic string, and then they will be used to construct TC and CT states for strings.  This will be done by using the tensor product of two Hilbert spaces for  string states.  Then  CT states will be used for the study of circuit complexity. 
\subsection{String  State}
Here, we shall briefly discuss the required bosonic string quantization framework for building up the corresponding string  state. The   Polyakov action, which describes  the world sheet of bosonic  string theory, can be written as 
\begin{eqnarray}
    \textbf{S} =   \frac{1}{4 \pi \alpha'} \int d\sigma ^2   \eta^{\alpha\beta} \partial_{\alpha}X_{\mu}  \partial_{\beta} X^{\mu}
\end{eqnarray}
Here  $\alpha'$ is   related to the  length  of string $l_s^2$. We can apply the 
  Neumann boundary conditions for open strings, and write the solution to the equations of motion obtained from the Polyakov action as   
\begin{eqnarray}
\label{sol}
    X^{\mu}\left(  \tau,\sigma\right)=  x^{\mu}+l_s \tau p^\mu+ il_s\sum_{m>0}\frac{1}{m}{{\alpha}}^{\mu}_m e^{-im\tau}cos(m\sigma )
    \end{eqnarray}
The  Polyakov action has  gauge degrees of freedom, so before   quantizing this theory, we have to remove these unphysical gauge degrees of freedom. Usually, the BRST formalism in      conformal gauge is used to quantize this theory     \cite{gauge}. Thus, after adding    a gauge fixing term in the conformal gauge, the corresponding ghost term is also added to  the  Polyakov action. The   sum of the  Polyakov action with this gauge fixing term and ghost term represents a new  effective action. This new effective action is  invariant under the BRST transformation. The    Noether charge corresponding to the BRST invariance of  the effective action is represented by the BRST charge $Q$.  The   physical states of the theory are defined to be the states which satisfy $Q |\rm{phy}> =0$. It has been demonstrated that the BRST quantization in conformal gauge is equivalent to light-cone quantization \cite{gauge1, gauge0}. Thus, it is possible to consistently analyze string theory in  light-cone gauge  \cite{gauge2, gauge4}. In fact,        in a light-cone gauge, the    physical sector decouples from   the ghost sector  of the theory \cite{gauge5, gauge12, gauge14}.  This decoupling is similar to the decoupling which occurs  in ordinary electrodynamics \cite{guage15}. Due to this decoupling, it is possible to use the light-cone gauge, and directly analyze the physical sector of the theory   \cite{gauge2, gauge4}. It is important to note that the light-cone gauge string theory is physically equivalent to the conformal field string theory. In fact, the    finite field-dependent BRST transformation can be used to relate the    generating functional for string theory  in light-cone gauge to  the  generating functional for string theory in  conformal gauge \cite{gauge8}. 

Thus, in   a light-cone gauge, we can directly remove the gauge   degrees of freedom, and analyze the physical string states. In the light-cone gauge, first, the target space  coordinates  are defined as  $\{ X^+, X^-\},[X^i]_{i=1}^{24} $, where  $ X^+= ( X^0+X^{25}),$ and $  X^-= ( X^0-X^{25})$. Then it is observed that   the $ X^+$ direction does not contain any string  oscillations, and   the  oscillation in the $X^-$ direction can be expressed in terms of other string  oscillations.   Therefore, in the light-cone gauge, we can only deal with the physical string  oscillations in $[X^i]_{i=1}^{24}$ directions, and these can be used to obtain information about the theory.   Thus, the  commutation algebras in light-cone gauge can be expressed as  $[{\alpha}_{m}^{i},{\alpha}_{-m}^{j}]=m\eta^{ij}$. In this scenario, it is conventional to identify the momenta of the center of mass with the zero mode frequency as $\alpha_0^{i} = {l_s}{}p^{i}/\sqrt{2}= \sqrt{2 \pi T} p^{i}$ with  $i =1\cdots 24$. It is straightforward to show that the negative and positive frequency modes ${\alpha}_{-m}^i,{\alpha}_{m}^i$ behave as raising and lowering operators, respectively, with the total number operator being ${N}= \textstyle \sum_{m>0}\sum_{i=1}^{24} m N_{i, m} = \sum_{m>0}\sum_{i=1}^{24} ({\alpha}_{-m}^i{\alpha}_{m}^i)$. Thus, the ground state $|0;p\rangle$ is  defined to be annihilated by the lowering operators and to be an eigenstate of the center of mass momenta, 
${\alpha}_m^i|0;p\rangle = 0, $ with $ m > 0,  $ and $
 {p}^{i}|0;p\rangle =p^{i}|0;p\rangle 
$.
Within the framework, the excited string states can now be built by acting the raising operator ${\alpha}_{-m}^i$ on the ground state of momentum $p$ as $\textstyle{{\alpha}_{-m}^i|{0;p}\rangle}$. By using the ground state raising operators $\alpha_{-m}^i$ for $m>0$, the generic string state can be created.  \cite{onumber} 
\begin{eqnarray}
|{N;p}\rangle &=&  \prod_{m\geq 1}\prod_{i=1}^{24} {\frac{1}{\sqrt{{m}^{N_{i, m}}  N_{i,m}!}}} {({{\alpha}}_{-m}^i)}^{N_{i, m}}|{0;p}\rangle \label{a}  
\end{eqnarray}
where $N_{i, m}$ is the occupation number of excitations of the $m$th mode in the transverse $i$ direction.  The energy of the string  goes like the level (number operator),
\begin{equation}
    N=\sum_{i=1}^{24}\sum_{m=1}^{\infty} mN_{i, m}
\end{equation}
It may be noted that unlike usual quantum mechanics, where such states are obtained using the Hamiltonian, here they are obtained using an alternative operator called the mass-squared operator ${M}^2$ \cite{wani}. This mass operator can, in turn, be related to the number operator $N$  using the mass-shell condition  \cite{gauge4}
\begin{equation} {M}^2 = p^\mu p_\mu =  \frac{1}{\alpha^{\prime}}\left(\sum_{m>0}\sum_{i=1}^{24} \alpha_{-m}^i. \alpha_{m}^i -{a}\right) = \frac{1}{\alpha^{\prime}}( N - a),
\end{equation} 
 It may be noted that the mass-shell condition related $M^2$ to $N$, and $N$ is in turn related to the occupation number using  ${N}=\textstyle\sum_{i=1}^{24}\sum_{m>0} m N_{i,m}$.  The value of the normal ordering constant for open strings is $a = 1$, and so the lowest state is a tachyonic scalar (with ${M}^2<0$).
 
  
Thus, even though there is a mathematical similarity at the level of commutation algebra with a  harmonic oscillator, the physics of this system described is very different \cite{4}. For simple quantum mechanical systems, like a harmonic oscillator,  different states are classified by their energies, and the circuit complexity is calculated with respect to those states. Here, the different modes of a string represent different  particles, and the circuit complexity is used to investigate the properties of this internal dynamics of string theory.   However, due to the mathematical similarity of the commutator algebra of string theory and a harmonic oscillator, standard methods can be used to mathematically analyze string theory \cite{wani}. 
By following the standard method of constructing the coherent states, we can define ${\alpha}_{m}|\varphi_{m}\rangle=\varphi_m |\varphi_{m}\rangle$ to obtain the coherent state of a single world-sheet mode in terms of $|k\rangle$ string states, the complete methodology can be seen in Appendix, 

\begin{equation}\label{SingleModeCoherent}
|\varphi^i _{m}\rangle =\sum_{n=0}^{\infty}  \frac{\varphi_m^{n} e^{{-\frac{|\varphi^i _m|^2}{2}}} }{\sqrt{k(k-m)(k-2m)...(m)}}  |k\rangle.
\end{equation}
where $k = nm$. 
By repeating this procedure for different modes, the total coherent state can be obtained via the following relation 
\begin{equation}\label{StringCoherent}
|\Phi\rangle =\prod_{m\geq1}\prod_{i=1}^{24} |\varphi^i _{m}\rangle.
\end{equation}
It may be noted that the same expression of the single-mode coherent state \eqref{SingleModeCoherent} can be obtained by operating the factorized displacement operator ${D}(\varphi^i _m)$ on the ground state $|0;p\rangle$
\begin{equation}
|\varphi^i _{m}\rangle = {D}(\varphi^i _m)|0;p\rangle, \quad \text{with} \quad {D}(\varphi^i _m)=\exp(\varphi^i _m{\alpha}^i_{-m}-{\varphi^{i*} _m{\alpha}^i_{m}}).
\end{equation} 
The time evolution of the coherent state \eqref{StringCoherent} is calculated as follows (with $k =nm$)
\begin{eqnarray}
|\Phi(t)\rangle &=& \prod_{m\geq1}\prod_{i=1}^{24} \sum_{n=0}^{\infty}  e^{-i\sqrt{\frac{{N}_{i,m}-1}{\alpha'}}t} \frac{\varphi_m^{n} e^{{-\frac{|\varphi^i_m|^2}{2}}} }{\sqrt{k(k-m)(k-2m)...(m)}}  |k\rangle \nonumber \\
&=& \prod_{m\geq1} \prod_{i=1}^{24}  \sum_{n=0}^{\infty}  \frac{\varphi_m^{n} e^{{-\frac{|\varphi^i _m|^2}{2}}} }{\sqrt{k(k-m)(k-2m)...(m)}} {\Big( 1- i{\sqrt\frac{N_{i,m}-1}{\alpha^{\prime}}}t\Big)} |k\rangle ,
\end{eqnarray}
which will be used later in our study.
It may be noted that the momentum of the string $p$ is related to the occupation number by  the mass-shell condition. Average of the mass-squared can be written as  $\langle \hat{M}^2\rangle = \langle \Phi|\hat{M}^2|\Phi\rangle$. String coherent state  $|\Phi\rangle$  can be expressed in terms of    $|k\rangle$ states, and are not the mass eigenstates as seen in Appendix.

 Thus, the total momentum of the coherent state can occur through various combinations of modes (which physically correspond to different particles). Hence, it is possible to have different occupation numbers for the same momentum $p$. However, each of those combinations has to satisfy the mass-shell conditions, as the mass-shell condition related the operator $M^2$ to $N$, which in turn is related to the occupation number as $ N_{i,m}$. 

{Now we would like to introduce the new set of states ${|{N;p,\Phi \rangle}}$ which will be used to construct  suitable thermal  states. These states will be   produced by displacement operator acting upon the energy eigenstates \cite{a} 
 \begin{equation}
    |{N;p,{\Phi}\rangle}\equiv {D{(\Phi)}}|{N;p}\rangle
 \end{equation}
 where the displacement operator is defined as  
 \begin{equation}
     {D{(\Phi)}}\equiv{\prod_{m\geq1}\prod_{i=1}^{24}{D{(\varphi_m^i)}}}\equiv{\prod_{m\geq1}\prod_{i=1}^{24}\exp{(\varphi^i\alpha^i_{-m}-\varphi^{*i}\alpha^i_m)}}
 \end{equation}
 These new sets of states are now constructed using the  operators, 
${b}^{i}_{-m}\equiv{D{(\varphi^i_m)}}{\alpha}_{-m}^{i}{{D}^{-1}{(\varphi^i_m)}}$. 
 By using Baker-Campbell-Hausdorff (BCH) formula, we can write for total strings
 ${b}^{i}_{-m}\equiv\big[\alpha_{-m}^{i}- m\phi^{*i}_m\big]
$. 
So in this case the  string state can be viewed as a ground state because of
${b}^{i}_{m}|{\varphi^i}_m\rangle = 0$, such that we have
\begin{eqnarray}
    {b}^{i}_{m}|{N_{i,m};p,{\varphi^i}_m}\rangle &=& \sqrt{(N_{i,m};p)}\Big|{((N_{i,m};p)-{m}), {\varphi^i}_m}\Big\rangle
\\ 
     {b}^{i}_{-m}|{N_{i,m};p,{\varphi^i}_m}\rangle &=& \sqrt{(N_{i,m};p)+{m}}\Big|((N_{i,m};p)+{m}), \varphi^i_m \Big\rangle
\end{eqnarray}
So that we can obtain our general string   state  as 
\begin{eqnarray}
     |{N;p,{\Phi}}\rangle &=& \prod_{m\geq1}\prod_{i=1}^{24} \Bigg[{\frac{({b}^{i}_{-m})^{N_{i,m}}}{\sqrt{{m}^{N_{i,m}}  N_{i,m}!}}}\Bigg]|{0;p,{\varphi^i}_m}\rangle
\end{eqnarray}
This is the expression for   string    state. This state is obtained by the action of these new operators $B=\prod_{m\geq1}\prod_{i=1}^{24}{b}^{i}_{-m}$ on the vacuum state $|{0;p,{\Phi}}\rangle$. 
\subsection{Thermal String  State}\label{subsec2.2}
\rhead{Thermal String State}
Finite temperature problems in thermal theories are built on the zero-temperature formulations of the associated theory. Therefore, before delving into the construction of the bosonic string TFD state, one has to define the vacuum. Let us start with the consideration that the zero-temperature vacuum consists of two components, the left vacuum $|0;p, R \rangle $ (living in a Hilbert space $\mathcal{H}^{(L^i)}$), and the right vacuum $|0;p, R \rangle  $ (living in another Hilbert space $\mathcal{H}^{(R^i)}$). Therefore, the composite space $\mathcal{H}$ in which the vacuum $|0;p, L \rangle   \otimes |0;p, R \rangle  $ lives are given by the tensor product of the two Hilbert spaces \cite{4}. In that case, we have to be cautious as the relations (2.30) and (2.31) are different from their paper} as 
\begin{equation}
\mathcal{H}=\prod_{i=1}^{24}\mathcal{H}^{(L^i)} \otimes \mathcal{H}^{(R^i)} .
\end{equation} 
The ladder operators ${{\alpha}^{L^i}_m},{{\alpha}^{L^i}_{-m}},{{\alpha}^{R^i}_m} ,{{\alpha}^{R^i}_{-m}}$ (with $i =1\dots 24$) of the corresponding system obey the following commutation relations
\begin{eqnarray}
&& [{\alpha}_{m}^{L^i},{\alpha}_{-m}^{L^i}]=[{\alpha}_{m}^{R^i},{\alpha}_{-m}^{R^i}]= m \label{Comm1} \\
&& [{\alpha}_{m}^{L^i},{\alpha}_{-m}^{R^i}]=[{\alpha}_{m}^{R^i},{\alpha}_{-m}^{L^i}]=0. \label{Comm2}
\end{eqnarray}
It should be noted that the vacuum state $|0;p, L \rangle   |0;p, R \rangle  $ defined in the Hilbert space $\mathcal{H}$ is a true vacuum and it is different from the usual thermal vacuum state, which is a sort of excited state. Let us now study the thermal bosonic open string in contact with the thermal reservoir at temperature $T={\beta}^{-1}$. It should be noted that the vacuum $|0;p, L\rangle$ of our theory is defined by 
\begin{equation}
{\alpha}_{m}^{L^i}|{0;p},L \rangle  ={\alpha}_{m}^{R^i}|{0;p},R \rangle   \quad\text{for} \quad m > 0,
\end{equation}
with the commutation relations defined by \eqref{Comm1} and \eqref{Comm2}. However, in order to construct the TFD state, we shall take unitary Bogoliubov transformations 
\begin{eqnarray}
{\alpha}^{L^i}_{m}\rightarrow {\alpha}^{L^i}_{m}(\beta) =  {U}_m(\beta) {\alpha}^{L^i}_{m} {U}_m^\dagger(\beta) =  \Big[ \cosh\theta_m(\beta){\alpha}^{L^i}_{m}-\sinh\theta_m(\beta){\alpha}^{{R^i}\dagger}_{m}\Big] \\ 
\label{bogol1}  {\alpha}^{R^i}_{m}  \rightarrow {a}^{R^i}_{m}(\beta) =   {U}_m(\beta) {\alpha}^{R^i}_{m} {U}_m^\dagger(\beta) =  \Big[ \cosh\theta_m(\beta){\alpha}^{R^i}_{m}-\sinh\theta_m(\beta){\alpha}^{{L^i}\dagger}_{m}\Big],
\label{bogol2}
\end{eqnarray}
where ${U}_m(\beta)$ is given by 
\begin{equation}\label{UOperator}
{U}_m(\beta)= \exp\left[{\theta_m{(\beta)}}({\alpha}^{L^i}_{-m}{{\alpha}^{R^i}_{-m}} - {\alpha}^{L^i}_{m}{\alpha}^{R^i}_{m})\right],
\end{equation}
such that the commutations relations \eqref{Comm1} and \eqref{Comm2} are transformed as
\begin{eqnarray}
&& [{\alpha}_{m}^{L^i}(\beta),{\alpha}^{L^i}_{-m}(\beta)]=[{\alpha}_{m}^{R^i}(\beta),{\alpha}^{R^i}_{-m}(\beta)]=m \\
&& [{\alpha}_{m}^{L^i}(\beta),{\alpha}^{R^i}_{-m}(\beta)]=[{\alpha}_{m}^{R^i}(\beta),{\alpha}^{L^i}_{-m}(\beta)]=0.
\end{eqnarray}
Here, $\theta_m(\beta)$ is a temperature-dependent real parameter defined by the following relations
\begin{eqnarray}\label{coshT} 
\cosh\theta_m(\beta)&=&\left\{1+f_B(\omega_m)\right\}^{1/2}=\prod_{i=1}^{24}\left[1-\exp\left\{\frac{-m\beta\sqrt{N_{i,m}}}{\sqrt{\alpha'}}\right\}\right]^{-1/2},
\\ \label{sinhT}
\sinh{\theta_{m}{(\beta)}} &=&{\{f_B(\omega_m)\}}^{1/2}=\prod_{i=1}^{24}\left[\exp\left\{\frac{m\beta\sqrt{N_{i,m}}}{\sqrt{\alpha'}}\right\} -1\right]^{-1/2},  \\ \tanh{\theta_m(\beta)} &=& \prod_{i=1}^{24}\exp\left\{\frac{-m\beta\sqrt{N_{i,m}}}{{2}\sqrt{\alpha'}}\right\},
\end{eqnarray}
where $f_B(\omega_m)$ is the Bose-Einstein distribution function, with $\omega_m$ being the frequency of the $m^\text{th}$ string mode. Now, one can define the single-string mode TFD state as
\begin{equation}
 {\alpha}^{L^i}_{m} (\beta)|\text{TFD}\rangle_m=0={\alpha}^{R^i}_{m} (\beta) |\text{TFD}\rangle_m,
\end{equation} 
or equivalently
\begin{equation}\label{TFD1}
|\text{TFD}\rangle_m = {U}_m(\beta)|0;p,L \rangle  |0;p,R \rangle  .
\end{equation}
By using the transformation operator from \eqref{UOperator}, the total string TFD state can be expressed as, (for $D= 26$)
\begin{equation}
|\text{TFD}\rangle_s= \prod_{m\geq1} \prod_{i =1}^{24}{\left[\frac{1}{\cosh{\theta_{m}{(\beta)}}}\right]}^{D-2}e^{\tanh{\theta_{m}{(\beta)}}}\exp\left({\alpha}^{L^i}_{-m}{\alpha}^{R^i}_{-m}\right)|0;p,L \rangle  |0;p,R \rangle  ,
\end{equation}
which can be rewritten as
\begin{equation}\label{TFD} 
|\text{TFD}\rangle_s=\prod_{m\geq1} \prod_{i =1}^{24}\left[1-\exp\left\{\frac{-m\beta\sqrt{N_{i,m}}}{\sqrt{\alpha'}}\right\}\right]^{\frac{D-2}{2}}\exp\left\{\frac{-m\beta\sqrt{N_{i,m}}}{{2}\sqrt{\alpha'}}\right\}|N_{i,m};p,L \rangle  |N_{i,m};p,R \rangle ,
\end{equation}
for which we have utilized the expression given in \eqref{coshT} and \eqref{sinhT}. The TFD state \eqref{TFD} is a general expression for the $m^\text{th}$ mode. However, since our aim is to study finite temperature problems, it will be worth analyzing the behavior of the state with respect to a local observer. If the local observer is on the left-hand side, we can calculate the reduced density matrix by tracing over the right-hand side degrees of freedom   
\begin{equation}
{\rho}^{L}=\prod_{m\geq1}\prod_{i=1}^{24}{\rho}_m^{L^i}=\prod_{m\geq1}\prod_{i=1}^{24}Tr_{\mathcal{H}^{R^i}}|{{TFD}\rangle_m}{\; \;}_m\langle{TFD}|,
\end{equation}
   which describes the reduced dynamics of the state with respect to the local observer. Recall that the TFD state has been built from the vacuum, which is a direct product of left and right zero-temperature states. Therefore, the time evolution of each side can be considered to be independent of each other. However, following closely with previous work done on a harmonic oscillator \cite{a}, we take the same time evolution for each side, and  write  it for the total TFD state \eqref{TFD}   as
\begin{eqnarray} 
   |{TFD(t)}\rangle_s &=&\prod_{m\geq1}\prod_{i =1}^{24} e^{-i \sqrt{\frac{N_{i,m}-1}{\alpha^{\prime}}}t}|TFD\rangle_s \nonumber  \\
&=& \prod_{m\geq1} \prod_{i =1}^{24}\Big[ e^{-it\sqrt{\frac{{N_{i,m}}-1}{\alpha'}}} {U}_m(\beta)|0;p,L \rangle  |0;p,R \rangle   \Big] \nonumber \\
&=& \prod_{m\geq1} \prod_{i =1}^{24}\Big[  e^{-it\sqrt{\frac{N_{i,m}-1}{\alpha'}}} {U}_m(\beta,t) |0;p,L \rangle  |0;p,R \rangle   \Big],
\end{eqnarray}
where ${U}_m(\beta,t)$ and ${U}_m(\beta)$ are unitarily equivalent, i.e.
\begin{equation}\label{UOpT}
{U}_m(\beta,t)=\prod_{i=1}^{24} e^{-it\sqrt{\frac{{N_{i,m}}-1}{\alpha'}}} {U}_m(\beta) e^{it\sqrt{\frac{{N_{i,m}}-1}{\alpha'}}}.
\end{equation}
Collecting the expression of ${U}_m(\beta)$ from \eqref{UOperator}, we rewrite \eqref{UOpT} as follows
\begin{equation}\label{UOpT1}
{U}_m{(\beta,t)}=\prod_{i=1}^{24}\exp\left[ \left(\theta{_{m}{(\beta)}} e^{-it\sqrt{\frac{N_{i,m}-1}{\alpha'}}} ({\alpha}^{L^i}_{-m}{{\alpha}^{R^i}_{-m}} - {\alpha}^{L^i}_{m}{\alpha}^{R^i}_{m}) e^{it\sqrt{\frac{N_{i,m}-1}{\alpha'}}}\right)\right].
\end{equation}
Now, we can use the BCH formula to calculate,
\begin{equation}\label{BCH1}
e^{-it\sqrt{\frac{N_{i,m}-1}{\alpha'}}} ({\alpha}^{L^i}_{-m}{{\alpha}^{R^i}_{-m}} - {\alpha}^{L^i}_{m}{\alpha}^{R^i}_{m}) e^{it\sqrt{\frac{N_{i,m}-1}{\alpha'}}} = e^{-it\sqrt{\frac{N_{i,m}-1}{\alpha'}}} {\alpha}^{L^i}_{-m}{{\alpha}^{R^i}_{-m}} - {\alpha}^{L^i}_{m}{\alpha}^{R^i}_{m} e^{-it\sqrt{\frac{N_{i,m}-1}{\alpha'}}}.
\end{equation}

The relation \eqref{BCH1} simplifies \eqref{UOpT1} as
\begin{equation}\label{UOpT2}
{U}_m{(\beta,t)}=\prod_{i =1}^{24}\exp\left[ \Big( \left\{Z_m(\beta) {\alpha}^{L^i}_{-m}{{\alpha}^{R^i}_{-m}} - Z_m^\ast(\beta) {\alpha}^{L^i}_{m}{\alpha}^{R^i}_{m}\right\} \Big)\right],
\end{equation}
where $Z_m(\beta) =\theta{_{m}{(\beta)}} e^{-it\sqrt{\frac{N_{i,m}-1}{\alpha'}}}$. We can rewrite \eqref{UOpT2} in a more compact Gaussian form by expressing the ladder operators in terms of the canonical variables ${\zeta}^{a}=\prod_m \prod_{i=1}^{24}{\zeta}^{a_m}_{L^i, R^i} $ with $ {\zeta}^{a_m}_{L^i, R^i} = ({X}_m^{L^i}, {X}_m^{R^i}, {P}_m^{L^i}, {P}_m^{R^i})$
\begin{eqnarray}
&& {\alpha}^{L^i}_{m}=\frac{1}{\sqrt{2}}  \left({X}^{L^i}_{m} + i{P}^{L^i}_{m}\right), \quad  {\alpha}^{R^i}_{m}=\frac{1}{\sqrt{2}}\left({X}^{R^i}_{m} + i{P}^{R^i}_{m}\right), \\
&& {\alpha}^{L^i}_{-m}=\frac{1}{\sqrt{2}}\left({X}^{L^i}_{m} - i{P}^{L^i}_{m}\right), \quad  {\alpha}^{R^i}_{-m}=\frac{1}{\sqrt{2}}\left({X}^{R^i}_{m} - i{P}^{R^i}_{m}\right)
\end{eqnarray}
also, we can write the unitary operator for total string modes  as, 
\begin{equation}\label{UOpT3}
{U}{(\beta,t)}=\prod_{m\geq1}\prod_{i=1}^{24}\exp\left[ -\frac{i}{2} K^{(0)}_{{a_m}{b_m}}{\zeta}^{a_m}_{L^i,R^i}{\zeta}^{b_m}_{L^i,R^i}
\right],
\end{equation}
where  $K^{(0)}_{{a_m}{b_m}} $ can be expressed as 
\begin{equation}
K^{(0)}_{{a_m}{b_m}} =\theta_m(\beta)
\left(
\begin{array}{c c c c}
0 & \sin\Big(\sqrt{\frac{N_{i,m}-1}{\alpha'}}t\Big) & 0 & \cos\Big(\sqrt{\frac{N_{i,m}-1}{\alpha'}}t\Big) \\
\sin\Big(\sqrt{\frac{N_{i,m}-1}{\alpha'}}t\Big) & 0 &  \cos\Big(\sqrt{\frac{N_{i,m}-1}{\alpha'}}t\Big) & 0\\
0 &  \cos\Big(\sqrt{\frac{N_{i,m}-1}{\alpha'}}t\Big) & 0 &  -\sin\Big(\sqrt{\frac{N_{i,m}-1}{\alpha'}}t\Big)\\
\cos\Big(\sqrt{\frac{N_{i,m}-1}{\alpha'}}t\Big) & 0 &  -\sin\Big(\sqrt{\frac{N_{i,m}-1}{\alpha'}}t\Big) & 0
\end{array}
\right).
\end{equation}
Therefore, it turns out that the TFD state is a Gaussian state along with an unimportant phase, and such a form will be useful for calculating the complexity in the covariance matrix approach as we will show later.
\section{String Coherent State in Thermal Field}
\lhead{String Coherent State in Thermal Field}
\rhead{}
We are now ready to study the string coherent state in the thermal field. Before constructing the string thermal coherent $|\text{TC}\rangle_s$ and coherent-thermal $|\text{CT}\rangle_s$ states, we shall introduce the coherent state first. By following the standard definition, we can write the string coherent state as
\begin{equation}
|\gamma;p,L \rangle   |\eta;p,R \rangle   = \prod_{m\geq 1}\prod_{i= 1}^ {24} {D}_m(\gamma, \eta) |0;p,L \rangle   |0;p,R \rangle 
\end{equation} 
where the displacement operator is given by 
\begin{equation}
{D}_m(\gamma, \eta)\equiv {D}_m^{L^i}(\gamma){D}_m^{R^i}(\eta)= \exp\left(\gamma {\alpha}^{L^i}_{-m}+\eta {\alpha}^{R^i}_{-m}-\gamma^\ast{\alpha}^{L^i}_{+m}-\eta^\ast {\alpha}^{R^i}_{+m}\right)
\end{equation} 
 with $\gamma$ and $\eta$ being complex parameters, which characterize the eigenvalues of the left- and right-hand-side annihilation operators respectively. The operation of ${D}_m(\gamma)$ and ${D}_m(\eta)$ on the left and right-hand side annihilation operators displaces them in the optical phase space by $\gamma$ and $\eta$, respectively
\begin{eqnarray}
&& {D}_m^{L^i}(\gamma){\alpha}^{L^i}_{m}{D}_m^{\dagger {L^i}}(\gamma)=  \Big[ {\alpha}^{L^i}_{m}-m\gamma \Big] ={b}^{L^i}_{m},\label{U1}\\
&& {D}_m^{R^i}(\eta){\alpha}^{R^i}_{m}{D}_m^{\dagger {R^i}}(\eta)=  \Big[ {\alpha}^{R^i}_{m}-m\eta \Big] = {b}^{R^i}_{m}. \label{U2}
\end{eqnarray} 
A similar rule is also applied to the creation operators. Given the relations \eqref{U1} and \eqref{U2}, it is straightforward to verify that the operator
\begin{equation}
{\mathcal{V}_m}(\beta)=\exp\left[{\theta_m{(\beta)}}({b}^{L^i}_{-m}{{b}^{R^i}_{-m}} - {b}^{L^i}_{m}{b}^{R^i}_{m})\right]
\end{equation}
is unitarily equivalent to ${U}_m(\beta)$ (defined in \eqref{UOperator}), i.e. ${\mathcal{V}_m}(\beta)= {D}_m(\gamma, \eta) {U}_m(\beta) {D}_m^\dagger(\gamma, \eta)$. 

\subsection{Coherent-Thermal String State}
\lhead{String Coherent State in Thermal Field}
\rhead{Coherent-Thermal String State}
The coherent-thermal state $|\text{CT}\rangle_s$ is built by operating the operator ${\mathcal{V}}(\beta)$ on the coherent state
\begin{equation}\label{CT}
|\text{CT}\rangle_s =  {\mathcal{V}}(\beta)|\gamma;p,L \rangle   |\eta;p,R \rangle  = {\mathcal{V}}(\beta) {D}(\gamma, \eta) |0;p,L \rangle   |0;p,R \rangle  .
\end{equation}
Since, ${\mathcal{V}}(\beta)$ and ${U}(\beta)$ are related by a unitary transformation, we can express \eqref{CT} as follows
\begin{equation}\label{CT1}
|\text{CT}\rangle_s = {D}(\gamma, \eta) {U}(\beta)|0;p,L \rangle   |0;p,R \rangle   =  {D}(\gamma, \eta) |\text{TFD}\rangle_s,
\end{equation}
where we have used the relation \eqref{TFD1}. Therefore, the coherent-thermal state $|\text{CT}\rangle_s$ can also be defined as a state where the vacuum is first thermalized and then displaced in the optical phase space. By using the exact expression of $|\text{TFD}\rangle_s$ from \eqref{TFD}, we can obtain the exact form of $|\text{CT}\rangle_s$ as given in the following
\begin{equation}
|\text{CT}\rangle_s=\prod_{m\geq 1} \prod_{i= 1}^ {24}\left[1-\exp\left\{\frac{-m\beta\sqrt{N_{i,m}}}{\sqrt{\alpha'}}\right\}\right]^{\frac{D-2}{2}}\exp\left\{\frac{-m\beta\sqrt{N_{i,m}}}{{2}\sqrt{\alpha'}}\right\}|N_{i,m};\gamma;p,L_i \rangle  |N_{i,m};\eta;p,R_i \rangle  ,
\end{equation}
Now, for a local observer situated on the left hand side, the reduced density matrix becomes
\begin{equation}
{\rho}^L_{CT}=\prod_m \prod_{i=1}^{24}{\rho}^{L^i}_{m}=\prod_m \prod_{i=1}^{24}Tr_{\mathcal{H}^{R^i}} |CT\rangle_m \  _m\langle CT|=Tr_{\mathcal{H}^R}|CT\rangle_s \  _s\langle CT|
\end{equation}
By following the same procedure as discussed in section \ref{subsec2.2}, we can obtain the time evolution of the coherent-thermal state as
\begin{eqnarray} 
|\text{CT}(t)\rangle_s
&=& e^{-i\sqrt{\frac{{N_{i,m}}-1}{\alpha'}}t}|\text{CT}\rangle_s \nonumber \\
&=& {D}(\gamma, \eta,t) e^{-i\sqrt{\frac{{N_{i,m}}-1}{\alpha'}}t} |\text{TFD}\rangle_s \nonumber \\
&=& {D}(\gamma, \eta,t) |\text{TFD}(t)\rangle_s,
\end{eqnarray}
where
\begin{equation}\label{TimeD}
{D}(\gamma, \eta,t) = e^{i\sqrt{\frac{{N_{i,m}}-1}{\alpha'}}t} {D}(\gamma, \eta) e^{i\sqrt{\frac{{N_{i,m}}-1}{\alpha'}}t}.
\end{equation}
We utilize the BCH formula to compute the following 
\begin{eqnarray}
e^{-it\sqrt{\frac{N_{i,m}-1}{\alpha'}}} {\alpha}^{L^i}_{-m} e^{-it\sqrt{\frac{N_{i,m}-1}{\alpha'}}}
&=&\exp\left[\frac{-i mt\sqrt{N_{i,m}}}{\sqrt{\alpha'}}\right] {\alpha}^{L^i}_{-m}, \label{start}\\  e^{-it\sqrt{\frac{N_{i,m}-1}{\alpha'}}} {\alpha}^{R^i}_{-m} e^{-it\sqrt{\frac{N_{i,m}-1}{\alpha'}}}
&=&\exp\left[\frac{-i mt\sqrt{N_{i,m}}}{\sqrt{\alpha'}}\right] {\alpha}^{R^i}_{-m},\\
e^{-it\sqrt{\frac{N_{i,m}-1}{\alpha'}}} {\alpha}^{L^i}_{m} e^{-it\sqrt{\frac{N_{i,m}-1}{\alpha'}}} &=& \exp\left[\frac{-i mt\sqrt{N_{i,m}}}{\sqrt{\alpha'}}\right] {\alpha}^{L^i}_{m}, \\
e^{-it\sqrt{\frac{N_{i,m}-1}{\alpha'}}} {\alpha}^{R^i}_{m} e^{-it\sqrt{\frac{N_{i,m}-1}{\alpha'}}} &=& \exp\left[\frac{-i mt\sqrt{N_{i,m}}}{\sqrt{\alpha'}}\right] {\alpha}^{R^i}_{m}
\end{eqnarray}
 This leads to the simplification of \eqref{TimeD}
\begin{eqnarray}\label{TimeD1}
{D}(\gamma, \eta;t)=\prod_{m\geq 1}\prod_{ i=1}^{24}  \exp\left[ \left\{\gamma e^{-it\sqrt{\frac{N_{i,m}-1}{\alpha'}}}{\alpha}^{L^i}_{-m} + \eta e^{-it\sqrt{\frac{N_{i,m}-1}{\alpha'}}}{\alpha}^{R^i}_{-m}
\right.\right. \nonumber \\ \left. \left. - \gamma^\ast e^{it\sqrt{\frac{N_{i,m}-1}{\alpha'}}}{\alpha}^{L^i}_{m} - \eta^\ast e^{it\sqrt{\frac{N_{i,m}-1}{\alpha'}}}{\alpha}^{R^i}_{m}\right\}\right].
\end{eqnarray}
Let us now express \eqref{TimeD1} in a more compact form in terms of the canonical variables ${\zeta}^{a_m}_{L^i,R^i}=  ({X}_m^{L^i}, {X}_m^{R^i}, {P}_m^{L^i},$ ${P}_m^{R^i})$ as
\begin{eqnarray}\label{DmOp}
{D}(\gamma, \eta;t)&=& \prod_{m\geq 1}\prod_{i=1}^{24} \exp\left[ -{i}\lambda_{a_m} {\zeta}^{a_m}\right]\nonumber\\
&=&\prod_{m\geq 1}\prod_{i=1}^{24} \exp\left[ -i(\lambda_{X_m^{L^i}}X_m^{L^i}+ \lambda_{X_m^{R^i}}X_m^{R^i}+ \lambda_{P_m^{L^i}} P_m^{L^i} + \lambda_{P_m^{R^i}} P_m^{R^i})\right]
\end{eqnarray}
where 
\begin{eqnarray}
&& \lambda_{{X}^{L^i}_{m}}= {\sqrt{2}}\left[\text{Re}[\gamma(L)] \sin\left(mt\sqrt{\frac{N_{i,m}}{\alpha'}}\right) - \text{Im}[\gamma(L)] \cos\left(mt\sqrt{\frac{N_{i,m}}{\alpha'}}\right)\right], \\
&& \lambda_{{P}^{L^i}_{m}}= {\sqrt{2}}\left[\text{Re}[\gamma(L)] \cos\left(mt\sqrt{\frac{N_{i,m}}{\alpha'}}\right) - \text{Im}[\gamma(L)] \sin\left(mt\sqrt{\frac{N_{i,m}}{\alpha'}}\right)\right], \\
&& \lambda_{{X}^{R^i}_{m}}= {\sqrt{2}}\left[\text{Re}[\eta(R)] \sin\left(mt\sqrt{\frac{N_{i,m}}{\alpha'}}\right) - \text{Im}[\eta(R)] \cos\left(mt\sqrt{\frac{N_{i,m}}{\alpha'}}\right)\right], \\
&& \lambda_{{P}^{R^i}_{m}}= {\sqrt{2}}\left[\text{Re}[\eta(R)] \cos\left(mt\sqrt{\frac{N_{i,m}}{\alpha'}}\right) - \text{Im}[\eta(R)] \sin\left(mt\sqrt{\frac{N_{i,m}}{\alpha'}}\right)\right].\label{end}
\end{eqnarray}
 where $ m\sqrt{N_{i,m}/\alpha'}$ is the frequency of the string for the $m^{\text{th}}$ mode. Thus, it is clear that the state $|\text{CT}\rangle_s$ is non-Gaussian.
 
\subsection{Thermal-Coherent String State}
\rhead{Thermal-Coherent String State}
In contrast to the coherent-thermal state defined in \eqref{CT1}, the thermal coherent state is defined by a state where the vacuum is first displaced in the optical phase space and then thermalized
\begin{equation}\label{TC}
|\text{TC}\rangle_m = {U}_m(\beta) {D}_m(\gamma, \eta) |0;p,L \rangle   |0;p,R \rangle  =  {U}_m(\beta) |\gamma;p,L \rangle   |\eta;p,R \rangle.
\end{equation}
Using the Bogoliubov transformations, it can be shown that the thermal-coherent states are the eigenstates of the thermal annihilation operators, i.e.
\begin{eqnarray}
{a}^{L^i}_{m}(\beta) |\text{TC}\rangle_m &=& \gamma |\text{TC}\rangle_m, \\
{a}^{R^i}_{m}(\beta) |\text{TC}\rangle_m &=& \eta |\text{TC}\rangle_m.
\end{eqnarray}
It indicates that the thermal-coherent state $|\text{TC}\rangle_m$ can be obtained by displacing the thermal vacuum state $|\text{TFD}\rangle_m$ in the optical phase space
\begin{equation}\label{TC1}
|\text{TC}\rangle_m = {D}_m(\gamma,\eta,\beta)|\text{TFD}\rangle_m.
\end{equation}
Here the displacement operator can be written as
\begin{eqnarray}
{D}_m(\gamma,\eta,\beta) &\equiv & {U}_m(\beta) {D}_m(\gamma,\eta) {U}_m^\dagger(\beta) \nonumber \\
&=&\exp \left[\left\{\gamma{a}^{L^i}_{-m}(\beta) + \eta{a}^{R^i}_{-m}(\beta) - \gamma^\ast{a}^{L^i}_{m}(\beta) - \eta^\ast{a}^{R^i}_{m}(\beta)\right\}\right] \nonumber \\
&=& \exp \left[\left\{\tilde{\gamma}(\beta){a}^{L^i}_{-m} + \tilde{\eta}(\beta){a}^{R^i}_{-m} - \tilde{\gamma}^\ast(\beta){a}^{L^i}_{m} - \tilde{\eta}^\ast(\beta){a}^{R^i}_{m}\right\}\right]\nonumber \\
&=& {D}_m(\tilde{\gamma},\tilde{\eta}), \label{parameter}
\end{eqnarray}
where $\tilde{\gamma}(\beta)$ and $\tilde{\eta}(\beta)$ are given by
\begin{eqnarray}
\tilde{\gamma}(\beta) &=&  \left\{\cosh \theta_m(\beta)\gamma + \sinh\theta_m(\beta)\eta^\ast\right\}, \label{Para1}\\
\tilde{\eta}(\beta) &=& \left\{\cosh \theta_m(\beta)\eta + \sinh\theta_m(\beta)\gamma^\ast\right\}. \label{Para2}
\end{eqnarray}
From \eqref{parameter}, it is obvious that the operators ${D}_m(\gamma,\eta,\beta)$ and ${D}_m(\tilde{\gamma},\tilde{\eta})$ are physically equivalent apart from a parameter transformation given by \eqref{Para1} and \eqref{Para2}. Thus, we can rewrite \eqref{TC1} as
\begin{eqnarray}
|\text{TC}(\gamma,\eta)\rangle_m &=& {D}_m(\tilde{\gamma},\tilde{\eta})|\text{TFD}\rangle_m = |\text{CT}(\tilde{\gamma},\tilde{\eta})\rangle_m,
\\
   |\text{TC}(\gamma,\eta)\rangle_s &=& \prod_{m\geq1}\prod_{i=1}^{24}{D}_m(\tilde{\gamma},\tilde{\eta})|\text{TFD}\rangle_m = \prod_{m\geq1}\prod_{i=1}^{24}|\text{CT}(\tilde{\gamma},\tilde{\eta})\rangle_m, 
\end{eqnarray}
which indicates that the coherent-thermal and thermal-coherent states are physically equivalent apart from their different parameter dependents. The parameters being related to the temperature, one can conclude that the states $|\text{CT}\rangle_s$ and $|\text{TC}\rangle_s$ apparently possess a similar eigenvalue space with different parameter dependence. The time-evolution of the the state $|\text{TC}\rangle_s$ is realized as,
\begin{eqnarray}
|\text{TC}(t)\rangle_s &=& e^{-it\sqrt{\frac{{N_{i,m}}-1}{\alpha'}}}|\text{TC}\rangle_s \nonumber \\
&=& {D}(\gamma, \eta,\beta,t) e^{-it\sqrt{\frac{{N_{i,m}}-1}{\alpha'}}} |\text{TFD}\rangle_s \nonumber \\
&=&{D}(\gamma, \eta,\beta,t) |\text{TFD}(t)\rangle_s,
\end{eqnarray}
where
\begin{eqnarray}
{D}(\gamma, \eta,\beta;t) &=& e^{-it\sqrt{\frac{{N_{i,m}}-1}{\alpha'}}} {D}(\gamma, \eta,\beta) e^{it\sqrt{\frac{{N_{i,m}}-1}{\alpha'}}} \nonumber \\
&=& e^{-it\sqrt{\frac{{N_{i,m}}-1}{\alpha'}}} {D}(\tilde{\gamma}, \tilde{\eta}) e^{it\sqrt{\frac{{N_{i,m}}-1}{\alpha'}}} \nonumber \\
&=& {D}(\tilde{\gamma}, \tilde{\eta}, t).
\end{eqnarray}
Thus, we can state 
\begin{equation}
|\text{TC}_s(\gamma,\eta,t)\rangle= |\text{CT}_s(\tilde{\gamma},\tilde{\eta},t)\rangle,
\end{equation}
which clearly suggests that the rest of the analysis from \eqref{start} to \eqref{end} is similar to that of $|\text{CT}\rangle_s$ apart from the transformation of the parameters $(\gamma,\eta)\rightarrow(\tilde{\gamma},\tilde{\eta})$, which we do not repeat here.
\section{Circuit Complexity} \label{section 4}
\rhead{}
\lhead{Circuit Complexity}
In this section, we will adopt the covariance matrix approach to calculate the circuit complexity for coherent-thermal $|\text{CT}_s\rangle$ and thermal-coherent $|\text{CT}_s\rangle$ string states. However, we shall compute it for the $|\text{CT}_s\rangle$ only, as the same analysis will be equally valid for $|\text{TC}_s\rangle$, which we have discussed in detail in the previous section. Remember that both $|\text{CT}_s\rangle$ and $|\text{TC}_s\rangle$ are non-Gaussian, and our aim is to study the circuit complexity for the non-Gaussian string state. Now, we can Consider the string  system with target space  coordinates and momentum as 
\begin{eqnarray} 
{\zeta}^{a} = \prod_{m\geq1}\prod_{i=1}^{24} {\zeta}^{a_m}_{L^i,R^i} = \prod_{m\geq1}\prod_{i=1}^{24} ({X}^{R^i}_{m}, {X}^{L^i}_{m}, {P}^{R^i}_{m}, {P}^{L^i}_{m}), \\    {\zeta}^{b}= \prod_{m\geq1}\prod_{i=1}^{24}
{\zeta}^{b_m}_{L^i,R^i} = \prod_{m\geq1}\prod_{i=1}^{24}\left(
\begin{array}{c}
{X}^{R^i}_{m}\\
{X}^{L^i}_{m}\\
{P}^{R^i}_{m}\\
{P}^{L^i}_{m}\\
\end{array}
\right), 
\end{eqnarray} 
The commutation relations of $m$-canonical world-sheet variables can be written as 
\begin{equation}
\left[{\zeta}^{a_m}_{L^i,R^i},{\zeta}^{b_m}_{L^i,R^i}\right] = i m {\Omega}^{{a_m}{b_m}}_{L^i,R^i},   \quad \text{where} \quad  {  {\Omega}}^{{a_m}{b_m}}_{L^i,R^i}=\left(
\begin{array}{c c}
0 & I\\
-I & 0\\
\end{array}
\right).
\end{equation}
Thus, the commutation relations for the target space can be written as 
\begin{eqnarray}
\left[{\zeta}^{a},{\zeta}^{b}\right] =  i\prod_{m\geq1}\prod_{i=1}^{24} m{{\Omega}}^{{a_m}{b_m}}_{L^i,R^i}. 
\end{eqnarray} 
Here, we have taken  a product of world-sheet coordinates  and momentum ${\zeta}^{a_m}_{L^i, R^i}$ to obtain the target space coordinates and momentum ${\zeta}^{a}$. 
In what follows, we shall suppress the operator symbol. Here, $ {{\Omega}}^{{a_m}{b_m}}_{L^i, R^i}$ is an anti-symmetric tensor, which can be utilized to raise the indices, for instance
\begin{equation}
 T_{...c_m a_m d_m...} \equiv {\Omega}^{{a_m}{b_m}}_{L^i,R^i}{T_{...{c_m}{b_m}{d_m}...}}.
\end{equation}

The one-point function for a pure Gaussian state is known to vanish. Let's consider an arbitrary string state $|\psi_m\rangle$ of string mode ``$m$", such that its symmetric two-point function may be expressed as \cite{a} 

\begin{equation}\label{OPF}
G^{{a}{b}}=\prod_{m\geq1}\prod_{i=1}^{24} G^{{a_m}{b_m}}_{L^i,R^i}=\prod_{m\geq1}\prod_{i=1}^{24}\frac{1}{2} \langle{ \psi_m}|\{\zeta^{a_m} _{L^i,R^i},\zeta^{b_m} _{L^i,R^i}\}|\psi_m\rangle,
\end{equation}
where  the symbol $\{\cdot,\cdot\}$ represents the anti-commutation relation. If an arbitrary string
state $|\psi_m\rangle$ is a pure Gaussian state with 
$\prod_m\prod_{i=1}^{24}\langle{ \psi_m}|\{\zeta^{a_m} _{L^i,R^i},\zeta^{b_m} _{L^i,R^i}\}|\psi_m\rangle=0$, then it is completely characterized by
its symmetric two-point function, often referred to as its (symmetric) covariance matrix method \cite{38}. However, as we can consider  states which    are non-Gaussian,  the corresponding one-point functions are non-vanishing \cite{a}.  Collecting the expressions from \eqref{UOpT3} and \eqref{DmOp}, we can write the time-evolution of the coherent-thermal state as
\begin{equation}\label{circuit1}
|\text{CT}(t)\rangle_s = \prod_{m\geq1}\prod_{i=1}^{24}\Bigg\{\exp\left[ -\frac{i}{2}\lambda_{a_m} \zeta^{a_m} _{L^i,R^i}\right] \exp\left[ -\frac{i}{2} K^{(0)}_{{a_m}{b_m}}\zeta^{a_m} _{L^i,R^i}\zeta^{b_m} _{L^i,R^i}\right]\Bigg\} |0;p,L \rangle  |0;p,R \rangle   .
\end{equation}
Here we have dropped the overall phase factor $\exp{(-i \sqrt{\frac{N_{i,m}-1}{\alpha^{\prime}}}t)}$ of strings. For generalizing the covariance matrix approach, we shall consider a general string reference state $|\psi_m\rangle_{\mathcal{R}}$ of string mode ``$m$", which can be Gaussian or non-Gaussian. The state \eqref{circuit1} is connected to the string reference state $|\psi_m\rangle_{\mathcal{R}}$ by two unitary transformations
\begin{eqnarray}
|\text{CT}(t)\rangle_s = \prod_{m\geq1}\prod_{i=1}^{24} \mathbb{U}_{m_1}({L,R}) \mathbb{U}_{m_2}({L^i,R^i}) |\psi_m\rangle_{\mathcal{R}}, &&
\mathbb{U}_{m_1}({L^i,R^i})=\exp\left[ -\frac{i}{2}\lambda_{a_m} \zeta^{a_m}_{L^i,R^i}\right], \nonumber\\
\mathbb{U}_{m_2}({L^i,R^i})=\exp\left[ -\frac{i}{2} K_{{a_m}{b_m}}\zeta^{a_m}_{L^i,R^i}\zeta^{b_m}_{L^i,R^i}\right]. &&
\end{eqnarray}
The first one, $\mathbb{U}_{m_1}({L^i, R^i})$ causes the $m$-translations of the canonical variables $\zeta^{a_m}_{L^i, R^i}$ for which the gaussianity of the reference state is broken. Thus, we need to modify the relation \eqref{OPF}. For this, let us introduce the following one-point function 
\begin{equation}
\varphi^a = \prod_{m\geq1}\prod_{i=1}^{24} \varphi^{a_m}_{L^i,R^i}\equiv\prod_{m\geq1}\prod_{i=1}^{24}{\langle\psi_m|\zeta^{a_m}_{L^i,R^i}}{|\psi_m\rangle},
\end{equation}
and study how the one-point functions $\prod_{m\geq1}\prod_{i=1}^{24} \varphi^{a_m}_{L^i,R^i}$ and the symmetric two-point functions $\prod_{m\geq1}\prod_{i=1}^{24}$ $G^{{a_m}{b_m}}_{L^i,R^i}$ behave under the $m^{th}$ unitary transformation $\mathbb{U}_{m_1}(L,R)\mathbb{U}_{m_2}({L^i,R^i})$.  Now we can  express $\mathbb{U}_1^\dagger \zeta^{a} \mathbb{U}_1 $ as
 {\begin{eqnarray}\label{U1Action}
\mathbb{U}_1^\dagger \zeta^{a} \mathbb{U}_1 &=& \prod_{m\geq1} \prod_{i=1}^{24}\mathbb{U}_{m_1}^\dagger({L^i,R^i}) \zeta^{a_{m_1}}_{L^i,R^i} \mathbb{U}_{m_1}({L^i,R^i}) \\ \nonumber 
&=& \left[\prod_{m \geq1}\prod_{i=1}^{24}e^ {i\lambda_{b_m}\zeta^{b_m}_{L^i,R^i}} \zeta^{a_m}_{{L^i,R^i}}e^{-i\lambda_{b_m} \zeta^{b_m}_{L^i,R^i}}\right] = \prod_{m\geq1}\prod_{i=1}^{24}(\zeta^{a_m}_{L^i,R^i} + m{\lambda}^{a_m}), 
\end{eqnarray}}
where we have used the Baker–Campbell–Hausdorff formula  along with the commutation relations
\begin{equation}
  \left[i\lambda_{b_m}\zeta^{b_m}_{L^i,R^i}, \zeta^{a_m}_{L^i,R^i}\right]=  m\lambda^{a_m},\quad{ \left[i\lambda_{b_m}\zeta^{b_m}_{L^i,R^i}, i\lambda_{b_n}\zeta^{b_n}\right]=0 } \quad \text{with}\quad  \lambda^{a_m} \equiv \Omega^{{a_m}{b_m}}_{L^i,R^i}{\lambda_{b_m}} .
\end{equation}
Here we will denote all the coordinates by $(X^1,\cdots X^N, P^1, \cdots P^N)= ({X}^{R^i}_{m}, {X}^{L^i}_{m}, {P}^{R^i}_{m}, {P}^{L^i}_{m})$, where $N $ is the index representing the combined effect of $m$ and $i$. 
Thus, the first unitary transformation creates a $ \mathbb{R}^{2N}$ translation \cite{a}.  A similar calculation for the action of other unitary $\mathbb{U}_{m_2}({L^i,R^i})$ on the canonical variables $\zeta^{a_{m_2}}_{L^i,R^i}$  leads to the following,
\begin{eqnarray}\label{U2Action}
U_2^\dagger \zeta^{a} U_2 &=& \prod_{m\geq1}\prod_{i=1}^{24} U_{m_2}^\dagger \zeta^{a_{m_2}}_{L^i,R^i} U_{m_2} \nonumber \\
 &=&  {\prod_{m\geq1}\prod_{i=1}^{24}\left[e^{i K_{b_m c_m} \zeta^{b_m}_{L^i,R^i} \zeta^{c_m}} \zeta^{a_m}_{L^i,R^i} e^{-i K_{b_m c_m} \zeta^{b_m}_{L^i,R^i} \zeta^{c_m}}\right]}\nonumber\\&=&
\prod_{m\geq1}\prod_{i=1}^{24} \chi ^{a_m}_{b_m}\zeta^{b_m}_{L^i,R^i}
\end{eqnarray}
where we have utilized the commutation relation
\begin{eqnarray}
  \left[i K_{{b_m}{c_m}}\zeta^{b_m}_{L^i,R^i}\zeta^{c_m}, \zeta^{a_m}_{L^i,R^i}\right] =  m K^{a_m}_{b_m}\zeta^{b_m}_{L^i,R^i} \nonumber\\
  \left[i K_{{b_m}{c_m}}\zeta^{b_m}_{L^i,R^i}\zeta^{c_m}, i K_{{b_n}{c_n}}\zeta^{b_n}\zeta^{c_n}\right]=0
\end{eqnarray}
where $K^{a_m}_{b_m}=\Omega^{{a_m}{c_m}} K_{{c_m}{b_m}}$, and  $\chi^a _b\equiv e^{K^a _b}$. Therefore, the second unitary transformation forms a Lie group $ Sp(2N, \mathbb{R})$, which plays a role similar to the rotation in the Minkowski space. Combining the results from \eqref{U1Action} and \eqref{U2Action}, we obtain the total $m$-transformation of the canonical variables
\begin{eqnarray}
(\mathbb{U}_1\mathbb{U}_2)^\dagger \zeta^{a} (\mathbb{U}_1\mathbb{U}_2)&=&\prod_{m\geq1}\prod_{i=1}^{24} (\mathbb{U}_{m_1}({L^i,R^i})\mathbb{U}_{m_2})^\dagger({L^i,R^i}) \zeta^{a_m}_{L^i,R^i} (\mathbb{U}_{m_1}({L^i,R^i})\mathbb{U}_{m_2}({L^i,R^i}))\nonumber\\
&=&\prod_{m\geq1}\prod_{i=1}^{24}\left(\chi^{a_m}_{b_m}\zeta^{b_m}_{L^i,R^i} + m\lambda^{a_m}\right),
\end{eqnarray}
which is similar to Poincar\'e transformation in the Minkowski space-time. For such groups, the  full group generated by $\mathbb{U}_{1}\mathbb{U}_{2}$ has a similar structure to that of the Poincar\'e group, given by the semi-direct product of $ \mathbb{R}^{2N}$ by the transformations  $ Sp(2N, \mathbb{R})$ \cite{a} 
\begin{equation}
\mathbb{R}^{2N}\rtimes Sp(2N, \mathbb{R}).
\end{equation} 
Now we observe that under the combined action of $\mathbb{U}_{m_1}({L^i,R^i})\mathbb{U}_{m_2}({L^i,R^i})$,  the one point function transforms as
\begin{equation}\label{trafo1}
\varphi^{\prime^a}=\prod_{m\geq1}\prod_{i=1}^{24}\varphi^{\prime^{a_m}}_{L^i,R^i}=\prod_{m\geq1}\prod_{i=1}^{24} \langle{\psi^{\prime}_m|\zeta^{a_m}_{L^i,R^i}|\psi^{\prime}_m\rangle} = \prod_{m\geq1}\prod_{i=1}^{24}\left( \chi^{a_m}_{b_m}\varphi^{b_m}_{L^i,R^i} + m\lambda^{a_m}\right) ,
\end{equation}
whereas under the same transformation, the symmetric two-point function transforms as
\begin{eqnarray}\label{trafo2}
G^{\prime a b}&=&\prod_{m\geq1}\prod_{i=1}^{24}G^{\prime a_m b_m}_{{L^i,R^i}} \nonumber \\
&=& \frac{1}{2}\prod_{m\geq1}\prod_{i=1}^{24}\langle{\psi^\prime}_m|\{{\zeta^{a_m}_{L^i,R^i},\zeta^{b_m}_{L^i,R^i}}\}|\psi^{\prime}_m\rangle \nonumber \\
&=& \frac{1}{2}\prod_{m\geq1}\prod_{i=1}^{24}\langle\psi_m|(U_{m_1}({L^i,R^i})U_{m_2})^{\dagger}({L^i,R^i})\{\zeta^{a_m}_{L^i,R^i},\zeta^{b_m}_{L^i,R^i}\}(U_{m_1}({L^i,R^i})U_{m_2}({L^i,R^i}))|\psi_m\rangle \nonumber \\
&=& \prod_{m\geq1}\prod_{i=1}^{24}\left(\chi^{a_m}_{c_m} G^{c_m d_m}_{L^i,R^i}\chi^{b_m}_{d_m} + m\chi^{a_m}_{c_m}\varphi^{c_m}_{L^i,R^i}\lambda^{b_m} + m\chi^{b_m}_{d_m}\varphi^{d_m}_{L^i,R^i}\lambda^{a_m} + m^2\lambda^{a_m}\lambda^{b_m}\right)
\end{eqnarray}
Now the  transformations \eqref{trafo1} and \eqref{trafo2} can be written as
\begin{eqnarray}
&& \varphi^{\prime}=\prod_{m\geq1}\prod_{i=1}^{24} \varphi_{m}^{{\prime }}({L^i,R^i})= \prod_{m\geq1}\prod_{i-1}^{24}\left(\varphi^{a_m}_{L^i,R^i} \chi_{a_m}^T + m\lambda^{a_m}\right)    \\
&& G^{\prime}=\prod_{m\geq1}\prod_{i=1}^{24} G_m^{\prime}({L^i,R^i}) = \prod_{m\geq1}\prod_{i=1}^{24}\left(\chi_{a_m}G_{a_m}({L^i,R^i})\chi_{a_m}^T +  m\lambda_{a_m}^T\varphi^{a_m}_{L^i,R^i}\chi_{a_m}^T \right. \nonumber \\  && \left. \,\,\,\,\,\,\,\,\,\,\,\,\,\,\, + m\varphi_{a_m}^T({L^i,R^i})\chi_{a_m}\lambda_{a_m} + m^2\lambda_{a_m}^T\lambda_{a_m}\right) \label{trafo3}
\end{eqnarray}
These transformations represent a general result for two non-Gaussian string states, and it reduces for single modes to   $ {\varphi^\prime}_m({L^i,R^i}) = \varphi_{m}({L^i,R^i}) = 0$,   for the product of string modes as (for the Gaussian string state case) 
\begin{eqnarray} 
\prod_m\prod_{i=1}^{24}{\varphi^\prime}_m({L^i,R^i}) &=& \prod_m\prod_{i=1}^{24}\varphi_{m}({L^i,R^i}) = 0, \\  \prod_m\prod_{i=1}^{24} G_m^{\prime}({L^i,R^i}) &=&\prod_{m}\prod_{i=1}^{24}\chi_m G_m({L^i,R^i})\chi_m^T. 
\end{eqnarray} 
 Having defined the transformation rules for the non-Gaussian state, we can now introduce the total string covariance matrix as 
\begin{equation}
\bar{G}=\prod_{m\geq1}\prod_{i=1}^{24}\left(
\begin{array}{c c}
G_m({L^i,R^i}) & m\varphi_m^T({L^i,R^i})\\
m\varphi_m({L^i,R^i}) & m^2\\
\end{array}
\right).
\end{equation}
With this, we can express the transformation \eqref{trafo3} in the following form
\begin{equation}
\bar{G}^\prime =     \prod_{m\geq1}\prod_{i=1}^{24} \mathcal{U}_m\bar{G}_m({L^i,R^i}) \mathcal{U}^T_m,
\end{equation}
where $\mathcal{U}_m$ is given by 
\begin{equation}
\mathcal{U}_m=\left(
\begin{array}{c c}
\chi_m & \lambda_m^T\\
0 & m\\
\end{array}
\right).
\end{equation}
The quantum circuit connecting the string target state $\prod_m \prod_{i=1}^{24}|\psi_m\rangle_T$ and the string reference state $\prod_m \prod_{i=1}^{24}|\psi_m\rangle_{\mathcal{R}}$ can be built by a series of action of $\prod_{m\geq1}\mathcal{U}_m$. However, it is not necessary that the unitary transformation $\prod_{m\geq1}\mathcal{U}_m$ be unique. There can be many such transformations that can connect the string reference state to the string target state and the corresponding geodesics may have different lengths. Our task is to find an optimal unitary transformation that determines the minimal geodesic in the group manifold. The length of the minimal geodesic is termed as the circuit complexity \cite{38}. For example, one can find the existence of a stabilizer subgroup $S$ for the string reference state, i.e. $\forall~ \prod_{m\geq1}\mathcal{U}_{m_S} \in S$ which results in  \cite{a}
\begin{eqnarray}
U_{S}|\Psi\rangle_{\mathcal{R}} =\prod_{m\geq1}\prod_{i=1}^{24} \mathcal{U}_{m_S}|\psi_{m}\rangle_{\mathcal{R}} &=& \prod_{m\geq1} \prod_{i=1}^{24}e^{ (\mathcal{B}_{m_S})}|\psi_m\rangle_{\mathcal{R}} \nonumber \\
&=& \prod_{m\geq1}\prod_{i=1}^{24}|\psi_{m}\rangle_{\mathcal{R}} =|\Psi \rangle_{\mathcal{R}},
\end{eqnarray}
where ${\mathcal{B}_{m_S}}$ is the generator of the sub-algebra $\mathcal{S}$. This implies that if the unitary $\prod_{m\geq1} e^{(\mathcal{A}_m)}$ (or a geodesic $\mathcal{Y}_m(\tau) =\prod_{m\geq1}e^{ (\tau\mathcal{A}_m)}$ in the group manifold) connects the reference state to the target state, there will be many unitaries that can achieve the  target state of the total string modes because of 
\begin{equation}
\big\{e^{(\mathcal{A})}e^{ (\mathcal{B}_S)}\big\}|\Psi\rangle_{\mathcal{R}}=
e^{\mathcal{A}}|\Psi\rangle_{\mathcal{R}} = |\Psi\rangle_T.
\end{equation}
The unitaries $\mathcal{U}_{S}$ involving factors in the stabilizer subgroup always add to the length of the geodesic in the space of circuits, while only trivially acting on the reference. That is  the reason for  restricting  the subspace of circuits transverse to the stabilizer subspace. In other words, the geodesics corresponding to the unitaries $\mathcal{U}_{S}$ that connects the reference state to the target state will not have the minimal length. Therefore, it is obvious that the optimal geodesic can not correspond to any unitary that belongs to the stabilizer subgroup. To find the optimal geodesic, we use the inner product on the Lie algebra of the transformation group \cite{a}
\begin{equation}
\langle{\mathcal{A}},{\mathcal{B}}\rangle=\prod_{m\geq1}\prod_{i=1}^{24}\langle{\mathcal{A}_m},{\mathcal{B}_m}\rangle_i= \prod_{m\geq1}\prod_{i=1}^{24} Tr({\mathcal{A}_m} \bar{G}_{m_{\mathcal{R}}}({L^i,R^i}){\mathcal{B}_m^T}\bar{g}_{m_\mathcal{R}}({L^i,R^i})),
\end{equation}
where ${\mathcal{A}},{\mathcal{B}}$ are total  generators for strings, $\bar{G}_{\mathcal{R}}$ is extended covariant metric associated with the string reference state and $\bar{g}_{\mathcal{R}}$ is its inverse matrix. Subsequently, we use the string horizontal subspace that is transverse to the stabilizer string sub-algebra,
\begin{equation}
   \mathcal{H}:=\prod_{m\geq1}\prod_{i=1}^{24}\{\mathcal{A}_m \in \mathcal{G}_m|\langle{\mathcal{A}_m},{\mathcal{B}_{m_S}}\rangle_i =0,\quad \forall \; \;{\mathcal{B}_{S}}\in S\}
\end{equation}
It is obvious that the horizontal generator $\mathcal{A}=\prod_{m\geq1}{\mathcal{A}_m} \in {\mathcal{H}}$ does not belong to the stabilizer  string sub-algebra. Therefore,  the optimal geodesic in our case will still be generated by a horizontal string generator i.e, $\mathcal{Y}_m(\tau) =\prod_{m\geq1}e^{{(\tau\mathcal{A}_m})}$. Thus the complexity of the string target state will be given by 
\begin{equation}
    \mathcal{C}\Big(\prod_{m\geq1}\prod_{i=1}^{24}|\psi^i_{m}\rangle_{\mathcal{R}} \rightarrow \prod_{m\geq1}\prod_{i=1}^{24}|\psi^i_{m}\rangle_T\Big)=\prod_{m\geq1}\prod_{i=1}^{24}\|\mathcal{A}_m\|_i.
\end{equation}
Therefore, evaluating the string complexity is equivalent to finding the horizontal generator $\mathcal{A}_m$ 
\begin{equation}
  |\Psi\rangle_T=\prod_{m\geq1}\prod_{i=1}^{24} |\psi^i_{m}\rangle_T= e^{\mathcal{A}} |\Psi\rangle_{\mathcal{R}}.
\end{equation}
Thus, we have explicitly obtained the circuit complexity of coherent-thermal  states $|CT\rangle_s$ in  bosonic string theory. This was  done using  the    covariance matrix approach. Here  we    generated the optimal geodesics by a horizontal string generator  $\mathcal{A}$ and  then  obtained  the circuit complexity using  the  length of the minimal geodesics in the group manifold. Similar results have been obtained for a simple quantum mechanical system \cite{4}, which is represented by a single harmonic oscillator. As string theory in light-cone gauges could be mathematically  represented as a number of harmonic oscillators, these results could be viewed as a mathematical generalization of the earlier results obtained for a simple quantum mechanical system. In fact, if we result in our results to a single mode $m =1$, and for a single dimension in target space say $i=1$, then the results of this paper are directly related to the results obtained for a simple quantum mechanical system \cite{4}. As the string theory is formed from multiple modes, it was expected that the circuit complexity of string theory has to be much larger than simple quantum systems. This has been explicitly demonstrated  in this paper. However, this mathematical  similarity occurs only at the level of commutation algebra, and the physics of the system discussed here is very different from the physics of a  harmonic oscillator \cite{4}.  
The complexity for scalar fields has also been recently obtained \cite{fields}. 
Here, again we can mathematically view string theory in light-cone gauge as a theory of multiple two-dimensional scalar fields $X^i$, with additional structure imposed on those fields.
Thus, after calculating the  circuit complexity of scalar field theory, it is natural to calculate the circuit complexity of string field theory. Here, we observe that circuit complexity for fermionic fields has also been studied \cite{ferm}. This has motivated the study of circuit complexity  super-symmetric field theories \cite{super}. Thus,   these works \cite{ferm, super} along with the results of this paper, can be used to obtain the circuit complexity of super-string theories.

\section{Conclusions}\label{sec5}
\rhead{Conclusions}
In this paper, we first  used   the Polyakov action, which describes the world sheet of bosonic string theory, to obtain  the string coherent state. 
After that, we introduce a new set of states to construct thermal states, which we call string TFD states.
We defined the zero temperature vacuum, which consisted of  one left vacuum and  one right vacuum. We constructed  left and right Hilbert spaces from these vacuum states. Then we defined the total Hilbert space as a tensor product of left and right Hilbert spaces.
We investigated the  thermal bosonic open strings in contact with thermal reservoirs.
After that,  we introduce the unitary Bogoliubov transformation to construct TFD states.
We also explicitly  calculate the density matrix by tracing over the right  degrees of freedom.
We construct the coherent thermal string state by operating the unitary Bogoliubov operator on the string coherent state and, then, calculate its time evolution.
Then we construct thermal coherent string states, where we first displace the vacuum in the optical phase and then thermalize it. 
We adopted a covariance matrix approach to calculate the circuit complexity for the coherent-thermal string state.   We  generate the optimal geodesics by a horizontal string generator and define the complexity by the length of the minimal geodesics in the group manifold. Thus,  the calculation of string circuit complexity was equivalent to finding a horizontal generator.   
\\
It is possible to use the results of this present paper, to analyze the circuit complexity for specific reference states. This can be done by investigating the circuit complexity when the  reference state is a Gaussian state like the TFD state or the vacuum state.  
It would be interesting to extend this work to superstring theories. This can be done by first constructing a   coherent state in a specific superstring theory. Then we can generalize the construction of such a coherent state to a thermal coherent state. This can be used to obtain the  circuit complexity for such  thermal states of  superstring theory. However, before analyzing the circuit complexity for such  thermal states  of  superstring theory, we need to analyze the circuit complexity for supersymmetric  field theories. It would also be interesting to analyze the effect of background fields on the circuit complexities. It would also be interesting to analyze the effect of T-duality on circuit complexities. It is expected that for bosonic string theory, the circuit complexity will be invariant under a T-duality.  It would be interesting to analyze the effect of T-duality and mirror symmetry on the circuit complexities of superstring theories. This work can  then be analyzed to such theories with background fields. 
\\
It would be interesting to analyze the circuit complexity of string theory in conformal gauges. It is known that string theory in conformal gauge is equivalent to string theory in 
light-cone gauge. However, unlike in light-cone gauges, where the ghost sector decouples from the physical sector of the theory, the effect of the ghost sector has to be explicitly studied in conformal gauges. This is usually done using the BRST formalism, so it would be interesting to analyze the circuit complexity for a bosonic string using the BRST formalism. The BRST charge can also be used to construct the cubic string field theory. It would be possible to use the cubic string field theory to investigate the circuit  complexity of a system of strings interacting with each other through a cubic string interaction. This can be done by generalizing the results of this paper to cubic string field theory. It is possible to use finite field-dependent BRST transformation to relate light-cone string theory to string theory in conformal gauge. Thus, we can use the finite field BRST transformations to obtain the circuit complexity in conformal gauges using the results of this paper. 
\\
It is possible  to construct quadrature operators for thermal string states. Then these quadrature operators can be used to analyze generalized string coherent states.   Such    quadrature operators for thermal string states can be obtained from   quadrature operators  for different thermal modes of a  string.  It would be interesting to construct and analyze such quadrature operators for both open and closed bosonic string states. This can then be used to obtain the circuit complexity for such states. We can also use these states to  perform quantum state tomography for thermal string states. It is expected that this can also be generalized to quadrature operators for string field theories and, then, use it to analyze circuit complexity for a system of interacting strings. It would be interesting to analyze the effect of dualities on the circuit complexity using these quadrature operators. 
\section*{Appendix}
\subsection*{Quantum Harmonic  Coherent states}
Here we will review the construction of coherent states in quantum mechanics  \cite{Qmech}.  We will use it to motivate similar construction in  string theory, in the next section.  
The states $|\alpha\rangle$ defined by: $a|\alpha\rangle = \alpha|\alpha\rangle$, with $\langle \alpha|\alpha\rangle = 1$, are
called coherent states.  
In the $|n\rangle$ basis the coherent state $|\alpha\rangle$ is written as:
\begin{equation}
    |\alpha\rangle= \sum_{n=0}^{\infty} c_n|n\rangle 
\end{equation}
Multiplying this expression from the left with the bra $\langle m|$ gives an expression for
the coefficients $c_m$, so:
\begin{equation}
    \langle m| \alpha\rangle=\sum_{n=0}^{\infty} c_n\langle m|n\rangle =c_m
\end{equation}
Here, we used the fact that the wavefunctions of the harmonic oscillator are
orthonormal. As a result, we obtain the following expression:
\begin{equation}\label{123}
    |\alpha\rangle=\sum_{n=0}^{\infty} |n\rangle\langle n|\alpha\rangle
\end{equation}
Using the expression
\begin{equation}\label{12}
    |n\rangle=\frac{1}{\sqrt{n!}}(a^{\dagger})^n |0\rangle
\end{equation}
for the wavefunction $|n\rangle$, we obtain 
\begin{equation}
    \langle n|\alpha\rangle= \frac{1}{\sqrt{n!}}\langle 0|a^n \alpha\rangle=\frac{\alpha^n}{\sqrt{n!}}\langle 0|\alpha\rangle
\end{equation}
Combining this with Eqn.\eqref{123}, we obtain
\begin{equation}\label{13}
    |\alpha\rangle=\langle0|\alpha\rangle \sum_{n=0}^{\infty} \frac{\alpha^n}{\sqrt{n!}}|n\rangle
\end{equation}
The constant factor $\langle0|\alpha\rangle$ must still be determined, which can be done using
normalization since the coherent state $\alpha\rangle$ has to be normalized 
\begin{equation}
    1=\langle \alpha|\alpha\rangle=\Big(\sum_{m=0}^{\infty}\langle \alpha|m\rangle \langle m|\Big)\Big(\sum_{n=0}^{\infty} |n\rangle \langle n|\alpha\rangle \Big)
\end{equation}
After rearranging the obtained expression, orthonormality can be used resulting in
\begin{equation}
    1=\langle \alpha|\alpha\rangle=\sum_{m=0}^{\infty}\sum_{n=0}^{\infty}\langle \alpha|m\rangle \langle m|n\rangle \langle n|\alpha\rangle=\sum_{n=0}^{\infty}
    \langle \alpha|n\rangle  \langle n|\alpha\rangle
\end{equation}
Now we can use Eqn.\eqref{12} for the functions $|n\rangle$, and this results in
\begin{equation}
   1=\langle \alpha|\alpha\rangle=\sum_{n=0}^{\infty}\frac{1}{n!} ({\alpha^{\star}})^n \alpha^n\langle \alpha|0\rangle \langle 0|\alpha\rangle=\sum_{n=0}^{\infty}\frac{1}{n!} |\alpha|^{2n}|\langle 0|\alpha\rangle|^2
\end{equation}
Here, the exponential function of $|\alpha|^2$
can be recognized:
\begin{equation}
  1=\langle \alpha|\alpha\rangle=  |\langle 0|\alpha\rangle|^2 e^{|\alpha|^{2}|}
\end{equation}

Solving for $|\langle 0|\alpha\rangle|$ we get;
\begin{equation}
    |\langle 0|\alpha\rangle|=e^{-\frac{1}{2}|\alpha|^2}
\end{equation}
Substituting this into Eqn.\eqref{13}, we obtain the final form:
\begin{equation}
    |\alpha\rangle=e^{-\frac{1}{2}|\alpha|^2} \sum_{n=0}^{\infty} \frac{\alpha^n}{\sqrt{n!}}|n\rangle
\end{equation}
The coherent state $|\alpha\rangle$ can be expressed in terms of the displacement operator
$D(\alpha)$, which is given by $D(\alpha)=\exp{(\alpha a^{\dagger} - \alpha^{\star}a)}$

Now use Eqn.\eqref{12} for $|n\rangle$,  gives:
\begin{equation}
    |\alpha\rangle=e^{-\frac{1}{2}|\alpha|^2} \sum_{n=0}^{\infty} \frac{\alpha^n}{{n!}}(a^{\dagger})^n|0\rangle=e^{-\frac{1}{2}|\alpha|^2} e^{\alpha a^{\dagger}}|0\rangle
\end{equation}
Which is the general expression for a coherent state of the Harmonic oscillator. Since the Energy of the harmonic oscillator is quantized and is given as;
\begin{equation}
    H= \hbar\omega\left(a^{\dagger}a +\frac{1}{2}\right)
\end{equation}
So,  the average energy can be written as; $\langle H \rangle=\langle \alpha |H|\alpha\rangle$, and express  coherent state  $|\alpha\rangle$   in terms of    $|n\rangle$ states as 
\begin{equation}
    \langle \alpha |H|\alpha\rangle= \hbar\omega \langle \alpha|a^{\dagger}a +\frac{1}{2}|\alpha\rangle=\hbar\omega\left(|\alpha|^2 + \frac{1}{2}\right)
\end{equation}
\subsection*{String Coherent State}
Following our work of \cite{wani}, we have defined the string  world-sheet mode coherent state as states which  most closely resemble  the behavior of classical string  oscillatory modes. So, the string  world-sheet mode coherent state $|\varphi_m\rangle$ have been expressed as an eigenstate of $\hat\alpha_m$, with eigenvalue $\varphi_m$
\begin{equation}
\label{STS}
\hat\alpha_m |\varphi_m\rangle = \varphi_m|\varphi_m\rangle    
\end{equation}
where        $ |\varphi_m\rangle$ satisfying  $ \langle\varphi_m|\varphi_m\rangle = 1 $. In order to have an explicit expression for  string coherent state $ |\varphi_m\rangle$,   we expand it in terms of  $ |k\rangle$ string states as 
\begin{equation}
\label{STS8}
|\varphi_m\rangle =  \sum_{k=0}^{\infty} |k\rangle \langle k|\varphi_m\rangle
\end{equation}
Here we can use $\langle k|\hat\alpha_m | \varphi_m\rangle = \varphi_m \langle k|\varphi_m\rangle$ to obtain 
$\langle k+m|\varphi_m\rangle = {\varphi_m} \langle k|\varphi_m\rangle   /{\sqrt{k+m}} $.  Replace $k$ with $k - m$, we  obtain 
$
\langle k|\varphi_m\rangle ={\varphi_m}\langle k-m|\varphi_m\rangle    /{\sqrt{k}} $, and by repeating this    process, we also obtain 
$
\langle k-m|\varphi_m\rangle = {\varphi_m} \langle k-2m|\varphi_m\rangle/{\sqrt{k-m}}$. Using these expressions, we can write 
$
\langle k|\varphi_m\rangle = {\varphi_m^2} \langle k-2m|\varphi_m\rangle/{\sqrt{k(k-m)}}
$. Now 
repeating this procedure $n$ times, we obtain a general expression for $\langle k|\varphi_m\rangle$ as 
\begin{equation}
\langle k|\varphi_m\rangle =\frac{\varphi_m^n}{\sqrt{k(k-m)(k-2m)...(k-(n-1)m)}} \langle k-nm|\varphi_m\rangle   
\end{equation}
Now for $k\geq m$, and    $k-nm =b$, with $b=(0,1,2...m-1), \, n = (0, 1, 2, 3...)$,   we can write the expression for $\langle k|\varphi_m\rangle$ as 
\begin{equation} 
\label{STS5}
\langle k|\varphi_m\rangle =\frac{\varphi_m^{n}}{\sqrt{k(k-m)(k-2m)...m)}} \langle b|\varphi_m\rangle    
\end{equation}

We can obtain the expression for  $ \langle b |\varphi_m\rangle $ using the string displacement operator $D(\varphi_m)$) and the  normalization condition. We define the string  world-sheet mode  displacement operator  as the operator which  generates a  string  world-sheet mode coherent state from the vacuum state, $|\varphi_m\rangle = D(\varphi_m) |0\rangle$. Thus, we can write the explicit expression for the string  world-sheet mode displacement operator as 
\begin{equation}
\hat{D}(\varphi_m)= e^{{-\frac{|\varphi_m|^2}{2}}} e^{{\varphi_m \hat\alpha_{-m}}} e^{{\varphi_m^* \hat\alpha_m}}
\end{equation} 
Now to obtain the expression for $ \langle b|\varphi_m\rangle $, we observe that 
\begin{align}
\langle b|\varphi_m\rangle &= \langle b|  D(\varphi_m) |0\rangle =e^{{-\frac{|\varphi_m|^2}{2}}} \langle b| e^{{\varphi_m \hat\alpha_{-m}}} e^{{\varphi_m^* \hat\alpha_m}}  |0\rangle
\nonumber  
\\ 
&=   e^{{-\frac{|\varphi_m|^2}{2}}} \langle b| ( 1+\varphi_m \hat\alpha_{-m}+....)  ( 1+\varphi_m^*\hat\alpha_{m}+...) |0\rangle   
\end{align}
Thus,   using $ \hat\alpha_{m} |0\rangle = 0$, we obtain 
$
\langle b|\varphi_m\rangle = e^{{-{|\varphi_m|^2}/{2}}}  \delta_{0,b}   
$, which vanishes for $b \neq 0$, and so, we can write 
\begin{equation}
\label{STS6}
\langle 0|\varphi_m\rangle = e^{{-\frac{|\varphi_m|^2}{2}}}    
\end{equation}
In the  general expression of $\langle k|\varphi_m\rangle$, we observe that $\langle b |\varphi_m\rangle =0$ for $b \neq 0$, and  we obtain a non-vanishing expression only for  $k=nm$. So, 
using the expression for $\langle b|\varphi_m\rangle$ in the general expression for $\langle k|\varphi_m\rangle$, we obtain 
\begin{equation}
\label{STS7}
\langle k|\varphi_m\rangle = \frac{\varphi_m^{n} e^{{-\frac{|\varphi_m|^2}{2}}} }{\sqrt{k(k-m)(k-2m)...(k-(n-1)m)}}    
\end{equation}
Using this expression for $\langle k|\varphi_m\rangle$ (with $k =nm$) in the general expression for string  world-sheet mode coherent states, we can write an explicit expression for a string  world-sheet mode coherent state as  
\begin{equation}\label{60}
|\varphi_m\rangle =\sum_{n=0}^{\infty}  \frac{\varphi_m^{n} e^{{-\frac{|\varphi_m|^2}{2}}} }{\sqrt{k(k-m)(k-2m)...(m)}}  |k\rangle
\end{equation} 
By repeating this procedure for different string modes, we can obtain string  world-sheet mode coherent states for different string  modes.
Using these string  world-sheet mode coherent states, a coherent state for strings in the target space  can be expressed as   
\begin{equation}
 |\Phi\rangle 
 =\prod_m |\varphi_m\rangle
\end{equation}

 Average of the mass-squared can be written as  $\langle \hat{M}^2\rangle = \langle \Phi|\hat{M}^2|\Phi\rangle$. String coherent state  $|\Phi\rangle$  can be expressed in terms of    $|k\rangle$ states as
\begin{eqnarray}
\langle \Phi|\hat{M}^2|\Phi\rangle &=& \prod_{i=1}^{24}\prod_{m\geq1}\langle \varphi_m^i|\frac{1}{\alpha^{\prime}}(\hat{N}_{i,m} -1)|\varphi_m^i\rangle\nonumber\\
&=&\prod_{i=1}^{24}\prod_{m\geq1}\sum_{k}\Big{\langle} k \Big{|}  \frac{(\varphi^i_m)^{\frac{k}{m}} e^{{-\frac{|{\varphi^i_m}|^2}{2}}} }{\sqrt{k(k-m)(k-2m)...(m)}}\Big|  \frac{1}{\alpha^{\prime}}(\hat{N}_m^i-1)\Big|\frac{(\varphi^i_m)^{\frac{k}{m}} e^{{-\frac{|{\varphi^i_m}|^2}{2}}} }{\sqrt{k(k-m)(k-2m)...(m)}}\Big| k\Big{\rangle}\nonumber\\
 &=&   \prod_{i=1}^{24}\prod_{m\geq1}\sum_{k}\frac{(\varphi^i_m)^{\frac{2k}{m}} e^{{-|{\varphi^i_m}|^2}}}{{k(k-m)(k-2m)...(m)}}\Big\{\frac{1}{\alpha^{\prime}}(k-1)\Big{\}}.  
\end{eqnarray}
As $i$ and $m$ increases, the value of average energy $\langle \Phi|\hat{M}^2|\Phi\rangle$   drops accordingly, because of the exponential term $e^{{-|{\varphi^i_m}|^2}}$. It may be noted here that the string coherent state $|\Phi\rangle$ is not a mass eigenstate.  Thus, we get a convergent result, which resembles what happens for coherent states of a harmonic oscillator in quantum mechanics. 
\\\\
\noindent \textbf{\large{Acknowledgments:}} S.\,D.\,acknowledges the support of research grants DST/INSPIRE/04/2016/001391 (by DST-INSPIRE, Govt. of India) and SB/SRS/2022-23/89/PS (by DST-SERB, Govt. of India).


\end{document}